\documentclass[a4paper,12pt]{article}

\newtheorem{remark}{Remark}

\usepackage{amsfonts}
\usepackage{bm}
\usepackage{graphicx}

\usepackage{amsmath}
\usepackage{amsfonts}

\usepackage{natbib}

\def\T{{ \mathrm{\scriptscriptstyle T} }}

\begin{document}

\date{}
\author{
D. Salmer\'on$^{1, 2}$, J. A. Cano$^3$ and C. P. Robert$^4$ \\
\textit{\small{$^1$CIBER Epidemiolog\'ia y Salud P\'ublica (CIBERESP), Spain.}}\\
\textit{\small{$^2$Servicio de Epidemiolog\'ia, Consejer\'ia de Sanidad y Pol\'itica Social,}}\\
\textit{\small{ Ronda de Levante 11, E30008-Murcia, Spain.}}\\
\textit{\small{$^3$Departamento de Estad\'istica e Investigaci\'on Operativa,}}\\ 
\textit{\small{Universidad de Murcia, E30100-Espinardo, Spain.}}\\
\textit{\small{$^4$PSL, Universit\'e Paris-Dauphine, CEREMADE, and}}\\
\textit{\small{CREST, 75775 Paris cedex 16, France.}}
}

\title{\textbf{Objective Bayesian hypothesis testing in binomial regression models with integral prior distributions}}

\maketitle

\begin{abstract}
In this work we apply the methodology of integral priors to handle Bayesian
model selection in  binomial regression  models with a general link function.
These models are very often used to investigate associations and risks in
epidemiological studies where one goal is to exhibit whether or not an exposure
is a risk factor for developing a certain disease; the purpose of the current
paper is to test the effect of specific exposure factors. We formulate the problem as a Bayesian model selection case and solve it using
objective Bayes factors. To construct the reference prior distributions on the
regression coefficients of the binomial regression  models, we rely on the
methodology of integral priors that is nearly automatic as it only requires the
specification of estimation reference priors and it does not depend on tuning
parameters or on hyperparameters within these priors. 
\newline


\end{abstract}

\section{Introduction}

In an epidemiological context the response variable is quite often
binary. Binomial regression models (and specially the
logistic regression model) are some of the main techniques on which
analytical epidemiology relies to estimate the effect of an exposure on
an outcome, adjusting for confounding. Other link functions can be
used: for example, when the objective is to model the ratio of
probabilities instead of the ratio of odds, the logistic
approximation can be inappropriate, see \cite{greenland2004}, and a
log-binomial model in which the link function is the logarithm is
preferable to a logistic model.

Binomial regression models make it possible to estimate the effect of several
risk factors and exposures on an outcome. While being able to estimate these
effects is paramount, the statistical validation of the underlying model is
equally of major importance. Due to this issue, epidemiological
studies most often associate to point estimations their associated confidence intervals and p-values for a contrast where the null hypothesis $H_0$ is the null effect of some specific factors of interest. However, a delicate issue is that the frequentist perspective makes it impossible to quantify the probability of the effect being true, that is, the probability of the alternative hypothesis $H_1$.

The purpose of the current work is to obtain the posterior probability of
the alternative hypothesis in a binomial regression model with a
general link function, using an automatic prior-modelling procedure that does not require the specification of tuning parameters or hyperpriors. Indeed, we
formulate here the hypothesis testing setting ($H_0$ versus $H_1$) as a model
selection problem and from a Bayesian perspective, since its expression is
based on the respective probabilities of both hypotheses after data are
observed. 

Each hypothesis provides a competing model to explain the
sample data. To set some notations, let us consider that,
under the null hypothesis the distribution of the
sample $y$ is $f_1(y\mid\theta_1)$, and under
the alternative one is $f_2(y\mid\theta_2)$. If both
models have a priori the same probability and the prior
distributions on the parameters are $\pi_i(\theta_i)$, $i=1,2$,
then the posterior probability of the alternative hypothesis  is
\begin{equation}
\frac{m_2(y)}{m_1(y)+m_2(y)}=\frac{B_{21}(y)}{1+B_{21}(y)},\label{prob_post_M2}
\end{equation}
where
\[
m_i(y)=\int
f_i(y\mid\theta_i)\pi_i(\theta_i)d\theta_i,\,\,\,i=1,2
\]
and $B_{21}(y)$ is the Bayes factor in favour of the
alternative hypothesis that is defined as
\[
B_{21}(y)=\frac{\int
f_2(y\mid\theta_2)\pi_2(\theta_2)d\theta_2}{\int
f_1(y\mid\theta_1)\pi_1(\theta_1)d\theta_1}.
\]

To compute the probability (\ref{prob_post_M2}) specification of the prior
distributions $\{\pi_1(\theta_1),\pi_2(\theta_2)\}$ on the parameters of the
models to be compared is previously needed.  In the literature are widely used
diffuse, vague or flat priors and objective ones like the Jeffreys prior
\citep{jeff1961} or the reference prior (Bernardo, \citeyear{bernardo1979}; Berger and Bernardo, \citeyear{berber1989}), to estimate the parameters of the regression models. However, the use of
these priors is not recommended for Bayesian model selection due to the fact,
among other reasons, that their formulation does not take into account the null
hypothesis, making difficult for $\pi_2(\theta_2)$ to be \textit{concentrated}
around that hypothesis, which is a widely accepted condition (see, e.g.,
\cite{2006Casella}, pages 157, 160). Another common problem with these
priors is that they are usually not proper, a property that leads to the
indetermination of the Bayes factor, although it is not the case for our
models, see \cite{ibrahimlaud1991} and \cite{chen2008}.

The literature on objective prior distributions for testing in
binomial regression models is quite limited. The intrinsic prior
distributions (Berger and Pericchi, \citeyear{BergerPericchi1996}; Moreno \textit{et al.}, \citeyear{moreno1998}) are objective priors which have been proved to behave well
in problems involving normal linear models, see \cite{2006Casella}; \cite{giron2006} and \cite{morenogiron2006_1}.
However, the implementation of this technique in binomial regression
models with a general link function has not been yet developed.
Recently \cite{Leon2011} have applied the intrinsic priors to the
problem of variable selection in the probit regression model. They
took advantage of intrinsic priors for normal regression models \citep{giron2006} due to the connection between the probit model
and the normal regression model with incomplete information.
Therefore their results only apply to probit models. Our setting is more
general in that it can be directly applied to other link functions like
the logit, the complementary log-log, the Cauchit and the probit
one. An extension of the Zellner's $g$-prior to
generalised linear models like binomial regression models has been
developed by \cite{sabanes2011}; however, this extension needs the
specification of the hyperprior distribution on the parameter $g$.

Our proposal here is to use integrals priors. This methodology
automatically provides prior distributions that do not depend on
hyperparameters, thus on values (or prior distributions) to be
subjectively assigned or estimated from the data, and it has proved
to perform satisfactorily in a number of situations, see \cite{canokesslersalmeron2007a}, (\citeyear{canokesslersalmeron2007b}) and \cite{canosalmeron2013}.

Next we formulate the problem. Suppose that
$\{(y_i,x_i);\,i=1,...,n\}$ are independent observations, where
$y_i$ is a Bernoulli distributed random variable, $y_i\sim
Ber(p_i)$, $x_i=(x_{i1},...,x_{ik})$ is a vector of covariates and
$X$ is the matrix with rows $x_1,...,x_n$. The probability $p_i$
is related with the vector $x_i$ through a link function such that
$g(p_i)=x_i\beta$, $i=1,...,n$, where
$\beta=(\beta_1,...,\beta_k)^{\T}\in\Theta\subseteq\mathbb{R}^k$ is
the vector of the regression coefficients and $x_{ik}=1$, that is
the intercept is $\beta_k$. For a given value
$k_0\in\{1,...,k-1\}$ we want to test the hypothesis 
\[
H_0: (\beta_1,...,\beta_{k_0})=(0,...,0)
\]
\emph{versus}
\[
H_1:(\beta_1,...,\beta_{k_0})\neq(0,...,0).
\]
This contrast is equivalent to the problem of selecting between
the models $M_1$ and $M_2$, with
\[
\begin{array}{ll}
M_1: & y_i\mid x_i,\theta_1\sim
Ber(p_i),\,g(p_i)=x_i\theta_1,\,i=1,...,n\\
& \theta_1=(\theta_{11},...,\theta_{1k})^{\T}\in\Theta_1\subseteq\mathbb{R}^k,\,\theta_{1j}=0, j=1,...,k_0,\\
\\
 M_2: & y_i\mid x_i,\theta_2\sim
Ber(p_i),\,g(p_i)=x_i\theta_2,\,i=1,...,n\\
&\theta_2=(\theta_{21},...,\theta_{2k})^{\T}\in\Theta_2\subseteq\mathbb{R}^k.
\end{array}
\]
There are $k-k_0$ unknown parameters in model $M_1$ and $k$ in
model $M_2$.

The probability of the alternative hypothesis after the sample $y$ is
observed is therefore the posterior probability (\ref{prob_post_M2}) of model $M_2$. The solution we propose here is to compute this posterior
probability based on a Bayes factor associated with integral priors.

\section{Integral Priors}

To compare the models $M_i:y\sim
f_i(y\mid\theta_i)$, $i=1,2$, and to build appropriate
objective priors, we rely on the integral
priors proposed in \cite{canokesslersalmeron2007a}, (\citeyear{canokesslersalmeron2007b}) and \cite{Cano2008integral}. Those priors 
are defined as the solutions $\{\pi_1(\theta_1),\pi_2(\theta_2)\}$
of the following system of two integral equations
\[
\pi_1(\theta_1)=\int\pi_1^N(\theta_1\mid z_1)m_2(z_1)dz_1
\]
and
\[
\pi_2(\theta_2)=\int\pi_2^N(\theta_2\mid z_2)m_1(z_2)dz_2,
\]
where $\pi_i^N(\theta_i)$ is an objective prior distribution used
for the purpose of estimation in model $M_i$,
\[
\pi_i^N(\theta_i\mid z)\propto
f_i(z\mid\theta_i)\pi_i^N(\theta_i),\,\,\, m_i(z)=\int
f_i(z\mid\theta_i)\pi_i(\theta_i)d\theta_i,\,\,\,i=1,2,
\]
and $z_1$
and $z_2$ are minimal imaginary training samples. See \cite{Cano2008integral} for details and motivations. While,
usually $z_1$ and $z_2$ are training samples
of a same size, this is not a requirement of the approach:
the constraint is to take $z_i$ of
minimal size under the constraint that $\pi_i^N(\theta_i\mid z_i)$
is a proper distribution.

The argument to derive these equations is that \textit{a priori}
the two models are equally valid and they are provided with ideal
unknown priors that yield to the true marginals, being \emph{a
priori} neutral for judging between both models. Moreover, these
equations balance each model with respect to the other one
since the prior $\pi _i(\theta _i)$ is derived from the marginal
$m_j(z_i),$ and therefore from $\pi_j(\theta_j)$, $i\neq j,$ as an
unknown expected posterior prior \citep{perezberger2002}.

Solving this system of integral equations is usually impossible. However,
there exists a numerical approach that provides simulations from those
integral priors. The above system of integral equations is indeed associated 
with a Markov chain with transition $\theta_2\rightarrow\theta_2^{\prime}$ that consists of the following four steps
\[
\begin{array}{ccc}
1.\,\,\, z_1 & \sim & f_2(z_1\mid\theta_2)\\

2.\,\,\, \theta_1 & \sim & \pi_1^N(\theta_1\mid z_1)\\

3.\,\,\, z_2 & \sim & f_1(z_2\mid\theta_1)\\

4.\,\,\, \theta_2^{\prime} & \sim & \pi_2^N(\theta_2^{\prime}\mid z_2).\\
\end{array}
\]
The invariant $\sigma$-finite measure associated with this Markov chain is the
integral prior $\pi_2(\theta_2)$. Therefore, it can be simulated indirectly by simulating this Markov chain provided the latter is recurrent.

In regression models, a training sample is associated with a set
of rows of the design matrix and therefore there exist different
training samples. To overcome this issue, in linear models, \cite{BergerPericchi2004} have suggested that imaginary training samples
can be defined as observations that arise by first randomly
drawing linearly independent rows from the design matrix and then
generating the corresponding observations from the regression
model. (A similar perspective is adopted in bootstrap.)

In the context of the integral priors methodology with regression models, 
this simulation of training samples can be easily adapted 
by first randomly drawing linearly independent rows of the design matrix and 
then generating the corresponding observations from the regression model in steps 1 and 3 of the above algorithm. 
This procedure is exactly how we proceed for binomial regression models.

Different training samples provide different amounts of
\textit{information} that can and do impact the resulting Bayes factor. In the
context of intrinsic priors, see \cite{BergerPericchi2004} about this issue. However, when using our procedure for integral priors, if a simulated training sample has a high \textit{information} amount in, say, step 1, it is \emph{compensated} in step 3 where a new training sample is drawn conditional on a new set of rows drawn independently of the previously rows used in step 1.
In addition, a pragmatic approach to the evaluation of integral priors is
to check whether or not they produce sensible and robust answers.

We stress that, for this model, the associated Markov chain is 
necessarily recurrent since the training samples have a finite state space and the full conditional densities $f_i(z\mid\theta_i)$, $i=1, 2$ are strictly
positive everywhere and therefore the Markov chain is irreducible and hence ergodic.

\section{Simulating imaginary training samples and posteriors: the theory}

To simulate Markov chains associated with the integral priors
two steps are required: first, we need to generate imaginary training
samples (steps 1 and 3) and second, we need to simulate from the
corresponding posteriors (steps 2 and 4). At this point we should
account for the fact that training samples are subsets of the data
such that the corresponding posteriors are proper. In the binomial regression problem,
if the vector $\tilde{y}=(\tilde{y}_1,...,\tilde{y}_k)$ is a subset of the data
and the submatrix $\tilde{X}$ with rows
$\tilde{x}_1,...,\tilde{x}_k$ of $X$ associated to $\tilde{y}$ is
of full rank, then the Jeffreys prior,
$\pi^N(\beta\mid\tilde{X})$, and the corresponding posterior,
$\pi^N(\beta\mid\tilde{y},\tilde{X})$, are proper distributions,
as can be seen in \cite{ibrahimlaud1991}. Furthermore, they stated that
this is the case for binary regression models, such as the
logistic, the probit and the complementary log-log regression
models. Therefore it is possible to select the imaginary training
samples $z_1$ and $z_2$ that are needed in steps 1 and 3 in such a way
that the dimensions of these samples are $k-k_0$ and $k$ respectively.
To generate these, we first have to select the
corresponding full rank submatrices $\tilde{X}$.

In addition, we need to simulate from the posterior
distribution $\pi^N(\beta\mid\tilde{y},\tilde{X})$. In binomial
regression models with link function $g$, it is usually the case
that  the posterior distribution of the regression coefficients does
not enjoy a simple and closed form, which complicates the simulation. 
We could consider an Accept-Reject algorithm based, for instance, on Laplace
approximations to the posterior distribution or use instead MCMC
steps. However, we propose a more efficient shortcut, namely that,
when $\tilde{y}$ has dimension $k$, $\tilde{y}_i\sim Ber(\tilde{p}_i)$, 
$g(\tilde{p}_i)=\tilde{x}_i\beta$, $i=1,...,k$, and the submatrix
$\tilde{X}$ above is of full rank, to simulate 
$\pi^N(\beta\mid\tilde{y},\tilde{X})$ is equivalent to simulate
$\pi^N(\tilde{p}_1,...,\tilde{p}_k\mid\tilde{y},\tilde{X})$ by
the change of variables
$\beta=\tilde{X}^{-1}(g(\tilde{p}_1),...,g(\tilde{p}_k))^{\T}$.
Usually $\Theta=\mathbb{R}^k$, although, when $\Theta$ reproduces
restrictions (\emph{e.g.} when $g(p)=\log(p)$), we can always repeat
simulations until the restriction is satisfied. The implementation
of this idea is straightforward since, whatever the link function $g$ is,
Jeffreys prior is
\[
\pi^N(\tilde{p}_1,...,\tilde{p}_k\mid\tilde{X})=\prod_{i=1}^k\frac{1}{\pi\sqrt{\tilde{p}_i(1-\tilde{p}_i)}}
\]
and therefore the posterior distribution,
\[
\pi^N(\tilde{p}_1,...,\tilde{p}_k\mid\tilde{y},\tilde{X})=
\prod_{i=1}^k\pi^N(\tilde{p}_i\mid\tilde{y},\tilde{X})=\prod_{i=1}^k
Beta(\tilde{p}_i\mid\tilde{y}_i+1/2,3/2-\tilde{y}_i),
\]
is easily  simulated. This shortcut is an important reason for choosing imaginary
training samples of appropriate and different sizes: $z_1$ of size $k_1=k-k_0$ and
$z_2$ of size $k$.

When working with intrinsic priors, 
\cite{CasellaMorenoJASA2009}, \cite{BergerPericchi2004}, \cite{consonni2011}, among others,
have found it more efficient to increase the size of the imaginary
training samples when the data come from a binomial distribution.
One way to achieve this in the case of binomial regression models,
while keeping the simplicity in simulating from the
posterior distribution of the regression coefficients, is to introduce
more than a single Bernoulli variable $\tilde{y}_i$ for each selected row
$\tilde{x}_i$. Concretely, if the vector
$\tilde{y}=(\tilde{y}_1,...,\tilde{y}_{k})$ is of dimension $qk$
($q$ being a positive integer),
$\tilde{y}_i=(\tilde{y}_i^1,...,\tilde{y}_i^q)$,
$\tilde{y}_i^t\sim Ber(\tilde{p}_i)$, $t=1,...,q$, and
$g(\tilde{p}_i)=\tilde{x}_i\beta$, $i=1,...,k$, then
$\pi^N(\tilde{p}_1,...,\tilde{p}_k\mid \tilde{y},\tilde{X})$ is
\[
\prod_{i=1}^k\pi^N(\tilde{p}_i\mid\tilde{y},\tilde{X})=\prod_{i=1}^k
Beta\left(\tilde{p}_i\mid
q\hat{y}_i+1/2,q\left(1-\hat{y}_i\right)+1/2\right),
\]
where $\hat{y}_i$ is the mean of the components of
$\tilde{y}_i$. As \cite{CasellaMorenoJASA2009} point out, the grade
of concentration about the null hypothesis is controlled by the
value of $q$. These authors apply this augmentation scheme to
independence in contingency tables, using intrinsic priors such
that the size of the imaginary training samples does not exceed
the size of the data. Taking advantage of this perspective, we propose that the number
of Bernoulli variables be a discrete uniform random variable
between $1$ and the number of times that each row is repeated in
the matrix $X$. If $N(x)$ is the number of times that the
row $x$ appears in the matrix $X$ and $q_i$ is a discrete uniform
random variable in $\{1,2,...,N(\tilde{x}_i)\}$, $i=1,...,k$, then
we can take $\tilde{y}=(\tilde{y}_1,...,\tilde{y}_{k})$,
$\tilde{y}_i=(\tilde{y}_i^1,...,\tilde{y}_i^{q_i})$,
$\tilde{y}_i^t\sim Ber(\tilde{p}_i)$, $t=1,...,q_i$, and
$g(\tilde{p}_i)=\tilde{x}_i\beta$, $i=1,...,k$. In this case the
posterior distribution
$\pi^N(\tilde{p}_1,...,\tilde{p}_k\mid\tilde{y},$
$\tilde{X},q_1,...,q_k)$ is
\[
\prod_{i=1}^k Beta\left(\tilde{p}_i\mid
q_i\hat{y}_i+1/2,q_i\left(1-\hat{y}_i\right)+1/2\right).
\]
\begin{remark}
The value $q_i\hat{y}_i$ can be directly generated from
the binomial distribution, avoiding the simulation of
$\tilde{y}_i^t$ at the end of steps 1 and 3, although no much
gain in execution time is derived from this choice.
\end{remark}
In the case of continuous covariates we need to only consider $N(x)=1$ 
since an increase in the size of the imaginary training samples
as described above makes no sense. When this happens, an alternative
could be to discretise the continuous covariates using quantiles and
to compute the value $N(x)$ for all the rows $x$ using the
discretised version instead of the continuous covariates, even though we work
later with the original matrix $X$.

\section{Running the Markov chain and computing the Bayes factor: the practice}

\subsection{Algorithm to run the Markov chain}

In this section, we describe in detail the algorithm used to simulate the Markov
chain with transition $\theta_2\rightarrow\theta_2^{\prime}$ that
is associated with our model selection problem. Recall that,
in order to simulate $z_1$ and $z_2$, we need to select full-ranked 
submatrices of $X$. The implementation is as follows: rows of $X$ are randomly
ordered and they are consecutively chosen until we have a full
rank matrix. The algorithm is divided in the following four steps:

\begin{itemize}
\item\textbf{Step 1}. Simulation of $z_1$.
\begin{description}
\item[-] Randomly select $k_1=k-k_0$ rows of the matrix $X$:
$\tilde{x}_1,...,\tilde{x}_{k_1}$, with the condition that if
$R_1$ is the submatrix of $X$ with these rows, and 
$R_2$ is the submatrix of $R_1$ with columns $k_0+1,...,k$,
then $\vert R_2\vert\neq 0$.

\item[-] Simulate $q_i\sim U\{1,...,N_1(\tilde{x}_i)\}$,
$i=1,...,k_1$, where $N_1(\tilde{x}_i)$ is the number of times
that the vector with the columns $k_0+1,...,k$ of $\tilde{x}_i$
appears in the design matrix of model $M_1$.

\item[-] Independently simulate $\tilde{y}_i^t\sim
Ber(g^{-1}(\tilde{x}_i\theta_2))$, $t=1,...,q_i$, $i=1,...,k_1$,
and take $z_1=(\tilde{y}_1,...,\tilde{y}_{k_1})$ where
$\tilde{y}_i=(\tilde{y}_i^1,...,\tilde{y}_i^{q_i})$.
\end{description}
\item\textbf{Step 2}. Simulation of $\theta_1$.
\begin{description}
\item[-] Simulate $\tilde{p}_i\sim Beta\left(\tilde{p}_i\mid
q_i\hat{y}_i+1/2,q_i\left(1-\hat{y}_i\right)+1/2\right)$,
$i=1,...,k_1$, and compute
\[
v=R_2^{-1}(g(\tilde{p}_1),...,g(\tilde{p}_{k_1}))^{\T}.
\]
\item[-] Take $\theta_1=(0,...,0,v^{\T})^{\T}$.
\end{description}

\item\textbf{Step 3}. Simulation of $z_2$.
\begin{description}
\item[-] Randomly select $k$ rows of the matrix $X$:
$\tilde{x}_1,...,\tilde{x}_{k}$, with the condition that if $S$ is
the submatrix of $X$ with these rows, then $\vert S\vert\neq 0$.

\item[-] Simulate $q_i\sim U\{1,...,N_2(\tilde{x}_i)\}$,
$i=1,...,k$, where $N_2(\tilde{x}_i)$ is the number of times that
$\tilde{x}_i$ appears in the design matrix of model $M_2$.

\item[-] Independently simulate $\tilde{y}_i^t\sim
Ber(g^{-1}(\tilde{x}_i\theta_1))$, $t=1,...,q_i$, $i=1,...,k$, and
take $z_2=(\tilde{y}_1,...,\tilde{y}_{k})$ where
$\tilde{y}_i=(\tilde{y}_i^1,...,\tilde{y}_i^{q_i})$.
\end{description}
\item\textbf{Step 4}. Simulation of $\theta_2^{\prime}$.
\begin{description}
\item[-] Simulate $\tilde{p}_i\sim Beta\left(\tilde{p}_i\mid
q_i\hat{y}_i+1/2,q_i\left(1-\hat{y}_i\right)+1/2\right)$,
$i=1,...,k$, and compute
\[
v=S^{-1}(g(\tilde{p}_1),...,g(\tilde{p}_{k}))^{\T}.
\]
\item[-] Take $\theta_2^{\prime}=v$.
\end{description}
\end{itemize}

\subsection{Computing the integral Bayes factor}

To compute the Bayes factor
\[
B_{21}(y)=\frac{\int
f_2(y\mid\theta_2)\pi_2(\theta_2)d\theta_2}{\int
f_1(y\mid\theta_1)\pi_1(\theta_1)d\theta_1}
\]
that is associated to the  integral priors
$\{\pi_1(\theta_1),\,\pi_2(\theta_2)\}$, and therefore to obtain
the posterior probability of model $M_2$ we can exploit the simulations
from both integral priors derived from
the Markov chain(s). Beginning with a value $\theta_2=\theta_2^0$,
each time the transition $\theta_2\rightarrow\theta_2^{\prime}$ is
simulated we obtain a value for $\theta_2$ and derive another one for
$\theta_1$. Therefore with this procedure we obtain two Markov
chains $(\theta_1^t)_t$ and $(\theta_2^t)_t$, whose stationary
probability distributions are respectively $\pi_1(\theta_1)$ and
$\pi_2(\theta_2)$. The ergodic theorem thus implies 
\[
\lim_{T\rightarrow\infty}\frac{\sum_{t=1}^T
f_2(y\mid\theta_2^t)} {\sum_{t=1}^T
f_1(y\mid\theta_1^t)}=B_{21}(y)
\]
and this result provides an inexpensive approximation to the Bayes factor
$B_{21}(y)$. However, the major difficulty with this approach is that when the
likelihood is much more concentrated than its corresponding integral prior, 
$\pi_i$, most of the simulations $\theta_i^t$ enjoy very small likelihood values, which
means that the approximation procedure is then inefficient, i.e.~results in a high variance. This
problem can be bypassed by importance sampling. However, importance sampling requires the ability
to numerically evaluate the integral priors, even though we are only able to
simulate from these distributions. The resolution of the difficulty is to
resort to nonparametric density estimations based on the Markov chains $(\theta_1^t)_t$ and
$(\theta_2^t)_t$. In the examples that we present here we have
used the kernel density estimation from the package np of R, see
\cite{racine2008_NP}. Concretely, if $\hat{\pi}_i(\theta_i)$ is
the kernel density estimation of $\pi_i(\theta_i)$, and
$G_i(\theta_i)$ is the importance density, then
\[
\int f_i(y\mid\theta_i)\pi_i(\theta_i)d\theta_i\approx\int\frac{f_i(y\mid\theta_i)\hat{\pi}_i(\theta_i)}{G_i(\theta_i)}G_i(\theta_i)d\theta_i.
\]
Then, simulating from $G_i(\theta_i)$ and evaluating
$f_i(y\mid\theta_i)$, $\hat{\pi}_i(\theta_i)$ and
$G_i(\theta_i)$, we can approximate the Bayes factor.

Alternatively, and still relying on kernel density estimation, the method of \cite{carlinchib1995} can be used to approximate the Bayes factor. A rough
comparison is provided by Laplace type approximations as in \cite{schwarz1978}. Closer to the original Rao-Blackwellisation argument of \cite{Gelfand1990},
the training sample provides the following Monte Carlo approximation
\[
\pi_i(\theta_i)\approx
\frac{1}{T}\sum_{t=1}^T\pi_i^N(\theta_i\mid z_j^t),\,j\neq i,
\]
where $z_j^t$ are simulations from $m_j(z)$, which is more accurate than a nonparametric estimation of the integral priors.
\section{Examples}

\subsection{Breast cancer mortality}

Table \ref{Greenland} reproduces a dataset on the relation of receptor
level and stage with the 5-year survival indicator, in a cohort of women with
breast cancer \citep{greenland2004}.

\begin{table}
\centering
\caption{Data relating receptor level and stage to 5-year breast
cancer mortality \citep{greenland2004}}{%
\begin{tabular}{ccccc}
\hline
  Stage & Receptor Level & Deaths &  & Total \\
\hline
1 & 1 & 2 & & 12\\
1 & 2 & 5 &  & 55\\
2 & 1 & 9 &  & 22\\
2 & 2 & 17 &  & 74\\
3 & 1 & 12 &  & 14\\
3 & 2 & 9 &  & 15\\
\hline
\end{tabular}}
\label{Greenland}

\end{table}

For this example we use the logistic link function. First, we compare the model with only the intercept and
the stage \emph{versus} the full model. A classical
logistic regression analysis finds an association between 
receptor level and mortality, with $2.51$ as the estimation for
the odds ratio and a p-value of $0.02$. 

In order to estimate the posterior probability of the full model $M_2$,
our importance sampling proposal is based on a normal distribution centred at
the maximum likelihood estimator $\hat{\theta}_i$ and covariance
$2\hat{V}_i$ where $\hat{V}_i$ is the estimated  covariance of
$\hat{\theta}_i$. We approximate $\pi_1(\theta_1)$ and
$\pi_2(\theta_2)$ based on the outcome of the Markov chain and
kernel density estimation as described in the previous section.
For the simulation times $T=1000$, $5000$ and $10,000$, we ran $50$
Markov chains of length $T$, while the importance sampling step also
relies on $T$ simulations. The mean and the standard
deviation of the $50$ estimations of the posterior probability of the model $M_2$
appears in Table \ref{Greenland_res1}, which indicates a high
probability of a true association between the receptor level and
mortality.


\begin{table}
\centering
\caption{Estimations of the posterior probability
 of the model $M_2$, based on $50$ Markov chains of length $T$ and 
 an importance sampling approximation supported by $T$ simulations}{%
\begin{tabular}{lcccccc}
\hline
 & & $T=1000$ & & $T=5000$ & & $T=10000$ \\
\hline
Mean & & $0.710$ & & $0.722$ & & $0.726$\\
Standard deviation & & $0.020$ & & $0.010$ & & $0.008$\\
\hline
\end{tabular}}
\label{Greenland_res1}

\end{table}


\begin{figure}[ht]
\centering
\includegraphics[scale=0.6]{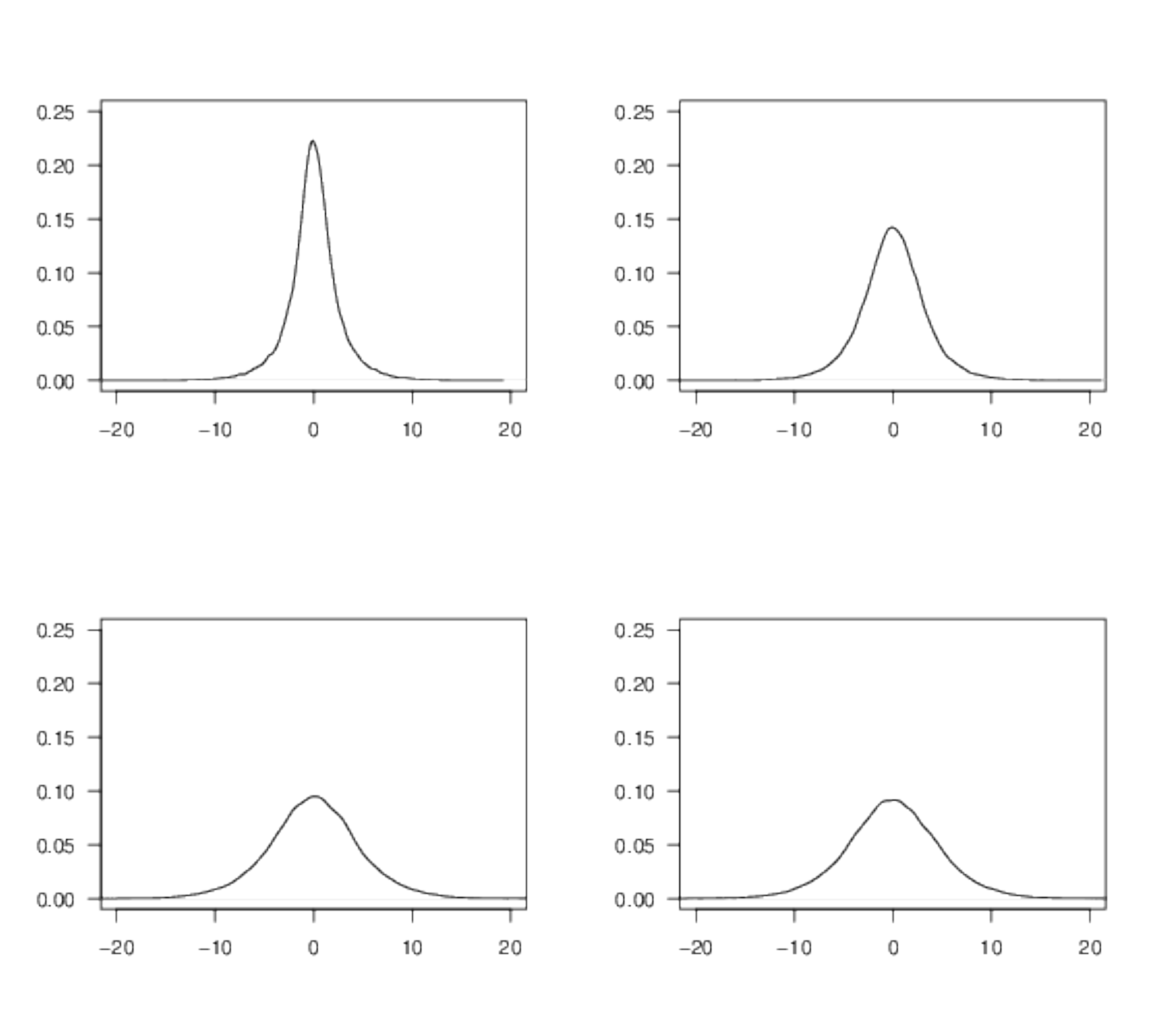}
\caption{Non-parametric approximations to the integral priors 
(top, left: receptor; top, right: intercept; bottom, left
and right: stage) based on 50,000 iterations of the associated Markov chain.}
 \label{figura_breastv3}
\end{figure}

Figure \ref{figura_breastv3} shows the integral priors for model
$M_2$. All the priors are centred around zero. In the first row there are the priors for the coefficient
of the receptor level and the intercept, the second row
corresponds to the stage. 

In this example with four regression coefficients and a sample of
size $192$, the high posterior probability of model $M_2$
indicates that there exists an association between mortality and
receptor level, although such probability is not conclusive.

On the other hand, it is well-known that stage is a factor
that is strongly related with mortality. We have computed the posterior
probability of the full model versus the model that includes the
intercept and the receptor level obtaining a posterior probability of $0.999$.
This very large value means that we can conclude that the most important predictor is by
far the stage if we are looking for a reduced model that
satisfactorily explain the data. For comparison, in this case the odds ratios are $3.11$ and $18.84$ and the p-values are $0.01485$ and
$5.34e-07$, respectively.

\subsection{Low birth weight}

The {\sf birthwt} dataset is made of 189 rows and 10 columns (see the object
{\sf birthwt} from the statistical software R). Data were collected at the
Baystate Medical Center, Springfield, Massachusetts in 1986 in order to attempt
to identify which factors contributed to an increased risk of low birth-weight
babies. Information was recorded from 189 women of whom 59 had low birth-weight
infants.  We use this dataset and the logistic link function to illustrate further the integral priors
methodology.

We first studied the association between the low birth-weight and smoking (two
levels), race (three levels), previous premature labours (two levels) and age
(five levels, defined by taking the intervals with included upper endpoints
$18, 20, 25, 30$ and $\infty$, respectively). We have considered as the reduced
model the one without the variable ``smoking".  The p-value associated with
the exclusion of ``smoking" is $0.014$ and the corresponding
estimation of the odds ratio is $2.62$.


\begin{figure}[ht]
\centering
\includegraphics[width=0.8\textwidth]{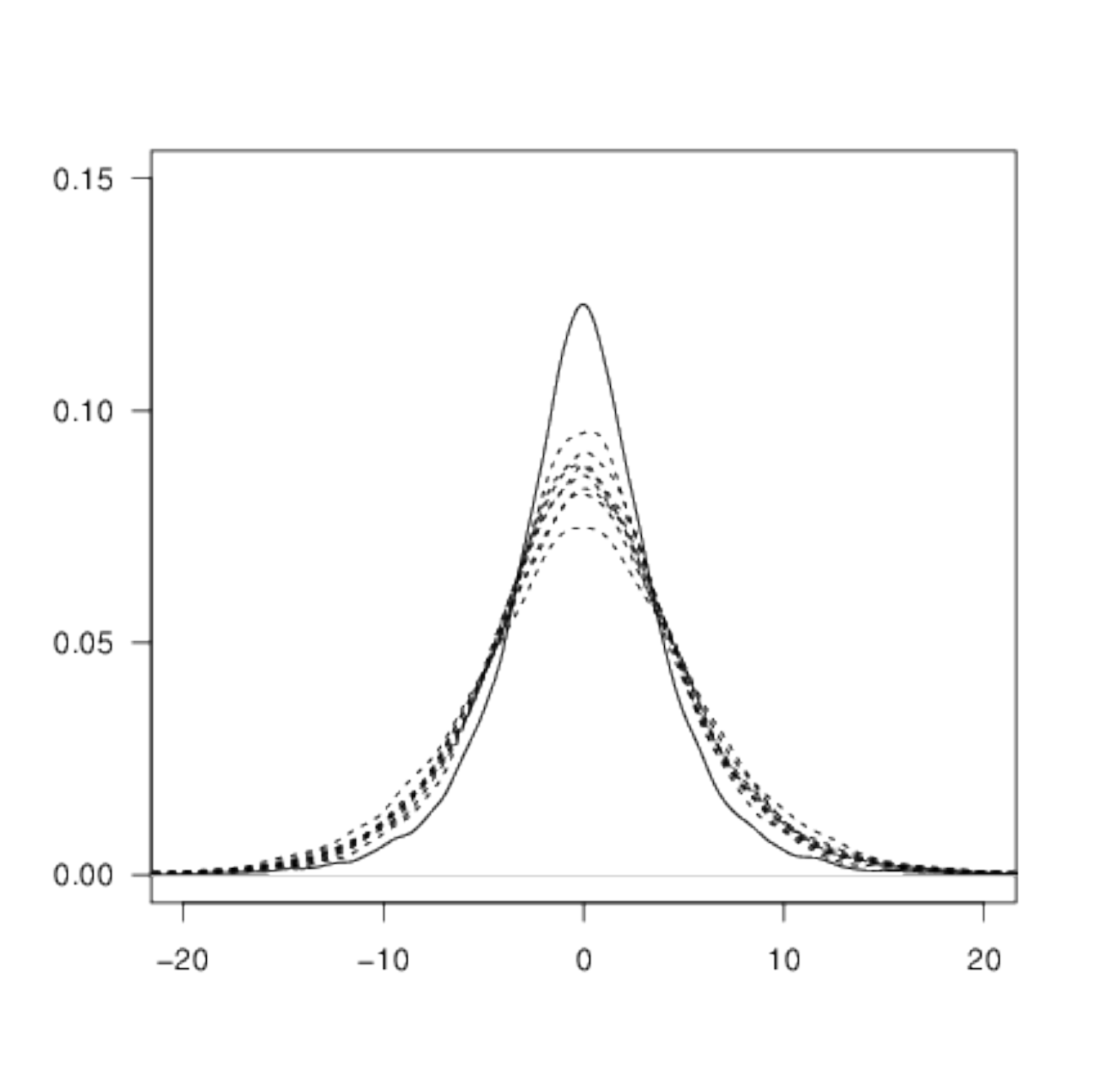}
\caption{Non-parametric approximations to the integral prior distributions for model $M_2$ associated with the
{\sf birthwt} dataset for the nine regression coefficients.}
\label{figura1}
\end{figure}

The analysis is based on $30,000$ iterations of the Markov
chain and $10,000$ simulations from the importance sampling density. It
yields $0.67$ as the posterior probability that smoking has an effect
over the low birth-weight. Figure \ref{figura1} produces an approximation
of the integral prior distributions for the nine regression coefficients. The
integral priors for all regression coefficients under model $M_2$
are very similar except the one for the smoking coefficient; this
prior is more concentrated about the null hypothesis. The standard
deviations for those priors are 4.2, 5.4, 5.5, 4.9, 5.7, 5.4, 5.8, 6.2 and 5.1,
respectively, showing again that the prior on the smoking
coefficient (first standard deviation) is more concentrated about
the null hypothesis while the others are similar.

To study the stability of these results, based on $T=10,000$, 
$20,000$ and $30,000$ iterations, we ran 30 Markov chains of length $T$
and, in parallel, importance sampling with $T$ simulations as well. Mean and standard deviation 
for the 30 estimations of the posterior probability of the model $M_2$ are reported in Table (\ref{tabla_ejemplo2}).
\begin{table}
\centering
\caption{Estimations of the posterior probability of the model $M_2$ running 30 Markov chains of length $T$ and 
importance sampling simulations based on $T$ simulations as well}{%
\begin{tabular}{lcccccc}
\hline
 & & $T=10000$ & & $T=20000$ & & $T=30000$ \\
\hline
Mean & & $0.671$ & & $0.673$ & & $0.681$\\
Standard deviation & & $0.0143$ & & $0.014$ & & $0.010$\\
\hline
\end{tabular}}
\label{tabla_ejemplo2}

\end{table}

\section{Conclusions}

Integral prior distributions have successfully been constructed towards
an objective Bayesian model selection analysis in binomial
regression models and the methodology has been applied in two
examples with the logistic regression. This analysis has been done
within the model selection framework and it remains completely
automatic since no other choice than the reference priors for the
competing models under consideration is requested. Although unrelated with
the purpose of this paper, this methodology can be applied to
variable selection problems, using an encompassing structure defined
from above or from below as done applying the
intrinsic priors methodology in \cite{Leon2011}. 

Furthermore, for the sake of comparison we have applied
the intrinsic prior methodology in \cite{Leon2011} to our examples. For the breast
cancer example we have calculated 30 times the posterior
probability of the full model using the package varSelectIP that
implements the intrinsic priors for the probit model, see
\cite{Leon2011}. The 30 computed values ranged
from 0.607 to 0.809 with a mean of 0.703 and standard deviation
0.055, thus exhibiting a similar answer but with more variability than the
integral methodology, see Table (\ref{Greenland_res1}). For the
second example (low birth-weight) the posterior probability of the
full model using the package varSelectIP 30 times ranged from
0.820 to 0.922 with a mean of 0.870 and standard deviation 0.024,
showing again that the integral methodology is more
stable that the one implemented with intrinsic priors (package
varSelectIP); at last, the conclusion using integral priors is
more conservative, which is a rather positive argument in medical studies when one is
trying to associate an exposure with an illness.


This feature could be the consequence of the property that despite
the fact that both the integral and the intrinsic priors are centred around
the null hypotheses, the corresponding null hypotheses are defined in different
ways since, when we use the intrinsic priors methodology developed in
\cite{Leon2011}, the intrinsic priors for all
models under consideration are centred around a null model where all
the $\beta's$ are zero except for the intercept,
that is the reference model for the intrinsic methodology.
Nevertheless, we should keep in mind that computations with
integral priors were made for the logistic model while those
for intrinsic priors were made for the probit model.

This work straightforward applies to other link functions
and can be extended to compare several link functions (non-nested
models). All the computations have been programmed in R and are freely
available at the web {\sf https://webs.um.es/dsm/miwiki/doku.php?id=investigacion}.

\section*{Acknowledgment}
This research was supported by the S\'eneca Foundation Programme
for the Generation of Excellence Scientific Knowledge under
Project 15220/PI/10. CPR was partly supported by Agence 
nationale de la recherche (ANR), on the project ANR-11-BS01-0010 Calibration.

\end{document}